# (15-crown-5)BiI$_3$ as a Building Block for Halogen Bonded Supramolecular Aggregates

Bettina Wagner, Johanna Heine*

Department of Chemistry and Material Sciences Center, Philipps-Universität Marburg, Hans-Meerwein-Straße, 35043 Marburg, Germany.

*Corresponding author, E-Mail: johanna.heine@chemie.uni-marburg.de

We present the synthesis and characterization of (15-crown-5)BiI$_3$ (**1**) and (15-crown-5)BiI$_3$·0.5TIE (**2**), a halogen bonded adduct with tetraiodoethylene (TIE), a typical halogen bond donor. Our results show that crown ether complexes of main group metal halides can be employed as halogen bond acceptors for the synthesis of new supramolecular aggregates and highlight the significant interaction between the two building blocks.

Halogen bonding has emerged as an important interaction in supramolecular chemistry,[1] and continues to be of great current interest regarding its theoretical description[2] and in fields of application from catalysis[3] to material science[4] and biochemistry.[5] Metal halide complexes can also be used in the construction of supramolecular aggregates involving halogen bonds, serendipitously or by design when halogenated solvents like halomethanes are used,[6,7] in many polyhalide compounds[8-15] or through addition of typical halogen bond donors like tetraiodoethylene (TIE) or 1,4-diiodotetrafluorobenzene.[16-19] A recent example by the Huber group has also shown that the activation of metal halide bonds through halogen bonding is important in homogeneous catalysis.[20]

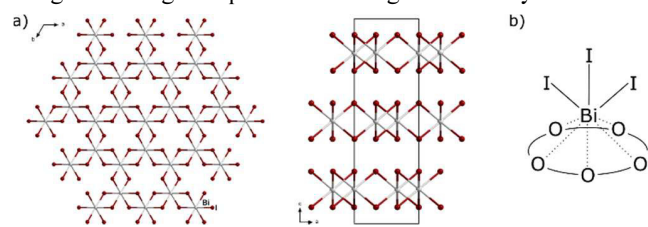

**Figure 1:** a) Crystal structure of BiI$_3$ along the *c* and *b* axis. b) Sketch of the chemical cut-out of a BiI$_3$-crown ether complex.

We wanted to use BiI$_3$-crown ether complexes as our building block for halogen bonded aggregates in an attempt to find suitable reagents to passivate the surface of BiI$_3$. Earlier reports on BiX$_3$-crown ether complexes [21,22] have shown that in most cases, the crown ether moiety occupied one hemisphere of the Bi$^{3+}$ ion, enforcing an arrangement of the halogen atoms that is fairly similar to that found on the (001) facet of BiI$_3$,[23] as illustrated in Figure 1. Additionally, an earlier work by Meyer and coworkers had provided a first example of such an aggregate in the form of (benzo-15-crown-5)BiI$_3$·I$_2$[24] and Bock has shown that halogen bonded aggregates with classic halogen bond donors like TIE could be obtained, for example as (18-crown-6)PbI$_2$·TIE.[25] Here, we present the synthesis, crystal structure and thermal and optical properties of (15-crown-5)BiI$_3$ (**1**) and the halogen bonded adduct (15-crown-5)BiI$_3$·0.5TIE (**2**).

(15-crown-5)BiI$_3$ (**1**) can be obtained by mixing stoichiometric amounts of 15-crown-5 and BiI$_3$ in solution. (15-crown-5)BiI$_3$·0.5TIE (**2**) can be obtained under similar conditions, but an excess of TIE has to be used to avoid the concomitant crystallization of **1** and **2**.

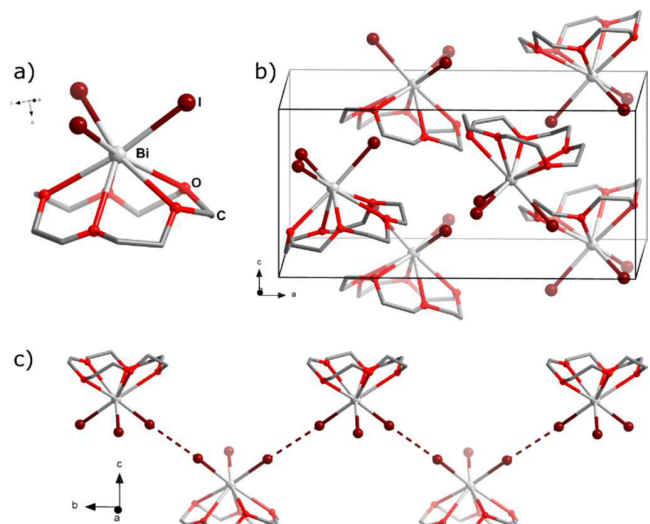

**Figure 2:** a) Molecular structure of **1**. b) Excerpt of the crystal structure of **1** highlighting the packing of the crown ether complexes. c) Iodine-iodine contacts (3.835 Å) in **1**, shown as fragmented lines. Hydrogen atoms omitted for clarity. Only one of the two positions of the disordered crown ether molecule shown.

(15-crown-5)BiI$_3$ (**1**) crystallizes in the orthorhombic space group $P2_12_12_1$ (No.19) as an inversion twin in the form of irregular orange crystals (see Figure S10). It is isostructural to both the chloride and the bromide analogue.[21,22] The molecular structure is shown in Figure 2a. The crown ether moiety is disordered over two positions (see Figure S1). Bi–I and Bi–O bond length (Bi–I: 2.889–2.906 Å; Bi–O: 2.825–3.048 Å) are in similar ranges as observed in (benzo-15-crown-5)BiI$_3$·I$_2$[24] (Bi–I: 2.873–2.964 Å; Bi–O: 2.824–2.911 Å). Bi–I bond length and angles (I–Bi–I: 89.5–90.6°) are also similar to those observed on the (001) facet of BiI$_3$ (Bi–I: 3.054 Å; I–Bi–I: 89.8°).[26] Some iodine-iodine contacts below the sum of the van der Waals radii[27] can be observed (Figure 2c), but are likely only enforced by the packing (Figure 2b), as the isostructural (15-crown-5)BiCl$_3$ features the same packing without short Cl···Cl contacts.[22]

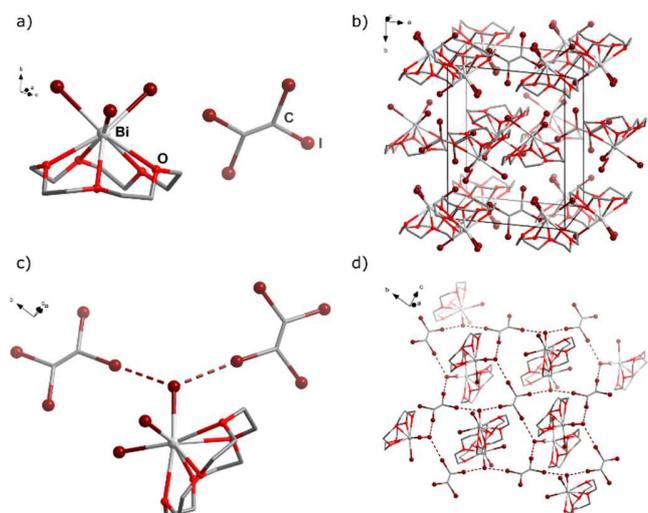

**Figure 3:** a) Molecular structure of **2**. b) Excerpt of the crystal structure of **2** highlighting the packing. c) Iodine-iodine contacts (3.526 and 3.646 Å) in **2**, shown as fragmented lines. C–I···I angles are 171.5° and 175.1°, Bi–I···I angles are 119.8° and 119.4°. While these values deviate from the ideal angles of 180° and 90°, the interaction can still be reasonably classified as a type II halogen bond. d) Fragment of the $4^4$-net created by the halogen bonded aggregate. Hydrogen atoms omitted for clarity.

(15-crown-5)BiI$_3$·0.5TIE (**2**) crystallizes in the monoclinic space group $P2_1/n$ (No.14) in the form of red blocks (see Figure S10) The molecular structure is shown in Figure 3, together with an excerpt of the crystal structure highlighting the packing, details of the halogen donor-acceptor interaction and the ensuing topology of the supramolecular aggregate. Bond lengths in the (15-crown-5)BiI$_3$ moiety (Bi–I: 2.907–2.948 Å; Bi–O: 2.751–3.011 Å) are similar to **1**. Bond length in the TIE molecule in **2** (C–I: 2.104–2.107 Å; C–C: 1.329 Å) are slightly longer than those found in the crystal structure of TIE (C–I: 2.088–2.097 Å; C–C: 1.337 Å),[28] well in line with expectations upon halogen bond formation. The contacts between the iodine atoms of the bismuth complex and those of the TIE molecule are displayed in Figure 3c, highlighting an arrangement that is typical for halogen bond donor-acceptor interactions.[29] As shown in Figure 3d, these interactions connect the two building blocks into a corrugated layer, that can be described as a $4^4$-square net where one of the iodine atoms of the (15-crown-5)BiI$_3$ moiety connects individual TIE molecules.

**1** and **2** show good thermal stability, decomposing at 260°C and 200°C, respectively (see Figures S5 and S6). IR- and Raman spectra of **1**, **2** and the starting material TIE (see Figures S8 and S9) indicate that the C–I-bond in the TIE molecule becomes weaker in **2**, similar to observations made for other halogen bonded aggregates involving TIE as a halogen bond donor.[30] We also investigated the influence of the adduct formation on the optical properties of our compounds. As shown in Figure 4, the onset of absorption is red-shifted by 20 nm for **2** in comparison to **1**. This can also be seen in photographs of the compounds (Figure 4, inset) and underscores a significant electronic interaction between the (15-crown-5)BiI$_3$ and TIE moieties.

In conclusion, our results show that crown ether complexes like (15-crown-5)BiI$_3$ (**1**) can act as halogen bond acceptors for typical halogen bond donors like TIE, allowing for the isolation of supramolecular aggregates like (15-crown-5)BiI$_3$·0.5TIE (**2**). We are now in the process of trying to translate our findings to the on-surface chemistry of exfoliated BiI$_3$ crystals. We also anticipate that other (crown ether)EX$_n$ (E = main group metal; X = halogen) complexes can be employed in a similar manner, opening up new possibilities for supramolecular chemistry, and that halogen bond donors like TIE can be used in place of I$_2$ to tune the properties of iodido metalate materials.

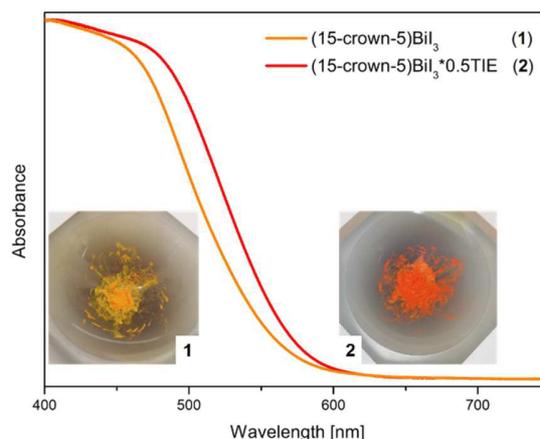

**Figure 4:** UV-Vis spectra of **1** and **2**, measured in diffuse reflectance, with photographs of the compounds shown as insets.

### Acknowledgments

We gratefully acknowledge support from the German Science Foundation (DFG) in the framework of the collaborative research center "Structure and Dynamics of Internal Interfaces" (SFB 1083). We thank Michael Hellwig for his help with EDX measurements, Martin Möbs for his help in obtaining IR- and Raman spectra and Bertram Peters for his help in obtaining µ-RFA data. J. H. thanks Prof. Stefanie Dehnen for her constant support.

**Supporting Information on**

# (15-crown-5)BiI$_3$ as a Building Block for Halogen-Bonded Supramolecular Aggregates

*Bettina Wagner, Johanna Heine\**

Department of Chemistry and Material Sciences Center, Philipps-Universität Marburg, Hans-Meerwein-Straße, 35043 Marburg, Germany.

\*E-Mail: johanna.heine@chemie.uni-marburg.de

## Table of Contents

## Synthesis

BiI$_3$ was synthesized from the elements according to a literature procedure.[1] All other reagents were used as received from commercial suppliers. All reactions were performed under aerobic conditions. Dichloromethane (DCM) and ethanol (EtOH) were flash-distilled prior to use. CHN analysis was carried out on an *Elementar* CHN-analyzer. Details on additional analysis methods can be found in the respective sections below.

**(15-crown-5)BiI$_3$ (1)**. BiI$_3$ (59 mg, 0.1 mmol) suspended in 40 mL of a solvent mixture (DCM:EtOH 1:1). 15-crown-5 (20 µL, 0.1 mmol) was added and the yellow suspension was heated to reflux. The solution slowly was cooled down over 2 hours and filtered. Within a few days, orange crystals of **1** formed, which were washed with 5 mL ethanol and dried in air. (Yield: 63 mg, 0.078 mmol, 78 %).

Data for **1**: Anal. Calcd for C$_{10}$H$_{20}$BiI$_3$O$_5$, (M = 809.96 g mol$^{-1}$): C 14.83, H 2.49 %. Found: C, 15.06; H 2.60 %.

**(15-crown-5)BiI$_3$·0.5TIE (2).** BiI$_3$ (59 mg, 0.1 mmol) and C$_2$I$_4$ (265 mg, 0.5 mmol) were suspended in 40 mL of a solvent mixture (DCM:EtOH 1:1). 15-crown-5 (20 µL, 0.1 mmol) was added and the yellow suspension was heated to reflux. The solution was slowly cooled down over 2 hours and filtered. Within a few days, red crystals of **2** formed, which were washed with 5 mL ethanol and dried in air. (Yield: 67 mg, 0.062 mmol, 62 %).

Data for **2**: Anal. Calcd for C$_{11}$H$_{20}$BiI$_5$O$_5$, (M = 1075.75 g mol$^{-1}$): C 12.28, H 1.87 %. Found: C, 12.31; H 1.86 %.

## Crystallographic Details

Single crystal X-ray determination was performed on a *Bruker Quest D8* diffractometer with microfocus MoKα radiation (λ = 0.71073).

**Table S1:** Crystallographic data for (15-crown-5)BiI$_3$ (**1**), CCDC 2035112.

| | |
|---|---|
| Empirical formula | C$_{10}$H$_{20}$BiI$_3$O$_5$ |
| Formula weight | 809.94 |
| Crystal system | orthorhombic |
| Space group | $P2_12_12_1$ |
| a/Å | 16.6924(10) |
| b/Å | 13.9898(8) |
| c/Å | 7.9935(4) |
| Volume/Å$^3$ | 1866.67(18) |
| Z | 4 |
| $\rho_{calc}$g/cm$^3$ | 2.882 |
| μ/mm$^{-1}$ | 14.417 |
| F(000) | 1448.0 |
| Crystal size/mm$^3$ | 0.183 × 0.129 × 0.035 |
| Absorption correction (T$_{min}$/T$_{max}$) | multi-scan (0.4492/ 0.7452) |
| 2Θ range for data collection/° | 4.88 to 50.646 |
| Index ranges | -20 ≤ h ≤ 20, -16 ≤ k ≤ 16, -9 ≤ l ≤ 9 |
| Reflections collected | 40514 |
| Independent reflections | 3407 [R$_{int}$ = 0.0723, R$_{sigma}$ = 0.0289] |
| Data/restraints/parameters | 3407/29/158 |
| Goodness-of-fit on F$^2$ | 1.086 |
| Final R indexes [I>=2σ (I)] | R$_1$ = 0.0316, wR$_2$ = 0.0587 |
| Final R indexes [all data] | R$_1$ = 0.0447, wR$_2$ = 0.0646 |
| Largest diff. peak/hole / e Å$^{-3}$ | 1.86/-2.05 |

**Details of crystal structure refinement:** Bi and I atoms were refined anisotropically. Hydrogen atoms were assigned to idealized geometric positions and included in structure factors calculations. The 15-crown-5 molecule was disordered over two positions with an occupancy of ½. Atoms in the disordered moiety were refined isotropically. To ensure a proper model of the crown ether, DFIX restraints were used on a number of carbon-carbon and carbon-oxygen distances. Overall, the crystal structure was refined as an inversion twin with a BASF value of 0.50276.

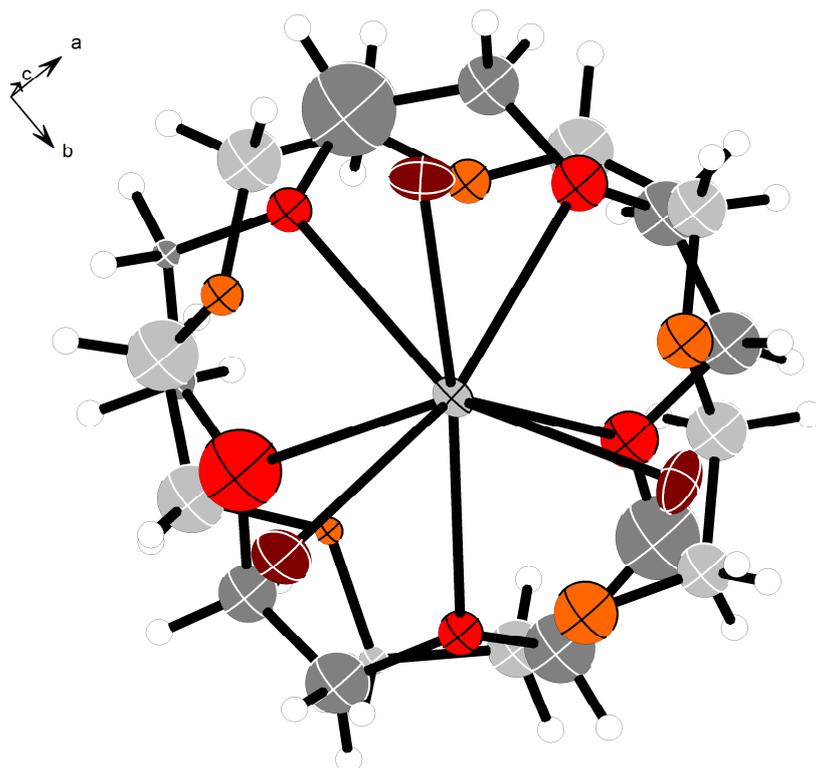

**Figure S1:** Asymmetric unit of **1**, ellipsoids at 50 % probability. The second position of the disordered 15-crown-5 moiety is shown in lighter colors.

**Table S2:** Crystallographic data for (15-crown-5)BiI$_3$·0.5TIE (**2**), CCDC 2035113.

| | |
|---|---|
| Empirical formula | C$_{11}$H$_{20}$BiI$_5$O$_5$ |
| Formula weight | 1075.75 |
| Crystal system | monoclinic |
| Space group | *P*2$_1$/*n* |
| a/Å | 12.8311(6) |
| b/Å | 14.6361(8) |
| c/Å | 13.2490(7) |
| β/° | 114.182(2) |
| Volume/Å$^3$ | 2269.8(2) |
| Z | 4 |
| ρ$_{calc}$g/cm$^3$ | 3.148 |
| μ/mm$^{-1}$ | 14.583 |
| F(000) | 1896.0 |
| Crystal size/mm$^3$ | 0.204 × 0.154 × 0.094 |
| Absorption correction (T$_{min}$/T$_{max}$) | multi-scan (0.3807/0.7452) |
| 2Θ range for data collection/° | 4.37 to 50.624 |
| Index ranges | -15 ≤ h ≤ 14, -17 ≤ k ≤ 17, -15 ≤ l ≤ 15 |
| Reflections collected | 61561 |
| Independent reflections | 4138 [R$_{int}$ = 0.0559, R$_{sigma}$ = 0.0208] |
| Data/restraints/parameters | 4138/0/199 |
| Goodness-of-fit on F$^2$ | 1.122 |
| Final R indexes [I>=2σ (I)] | R$_1$ = 0.0187, wR$_2$ = 0.0455 |
| Final R indexes [all data] | R$_1$ = 0.0210, wR$_2$ = 0.0467 |
| Largest diff. peak/hole / e Å$^{-3}$ | 1.73/-1.55 |

**Details of crystal structure refinement:** All non-hydrogen atoms were refined anisotropically. Hydrogen atoms were assigned to idealized geometric positions and included in structure factors calculations.

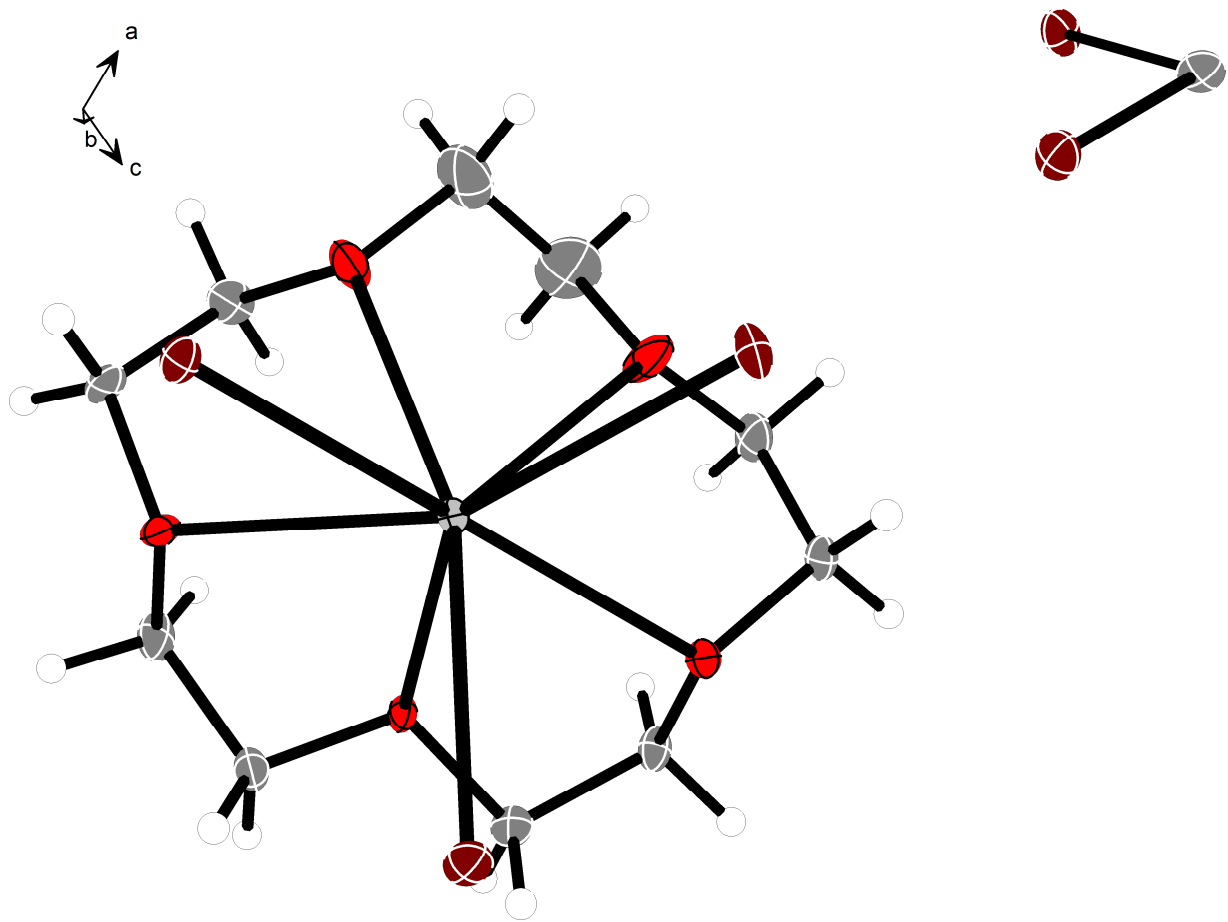

**Figure S2:** Asymmetric unit of **2**, ellipsoids at 50 % probability.

## Powder Diffraction

Powder patterns were recorded on a *STADI MP* (*STOE* Darmstadt) powder diffractometer, with CuKα1 radiation with λ = 1.54056 Å at room temperature in transmission mode. The patterns confirm the presence of the respective phase determined by SCXRD measurements and the absence of any major crystalline by-products.

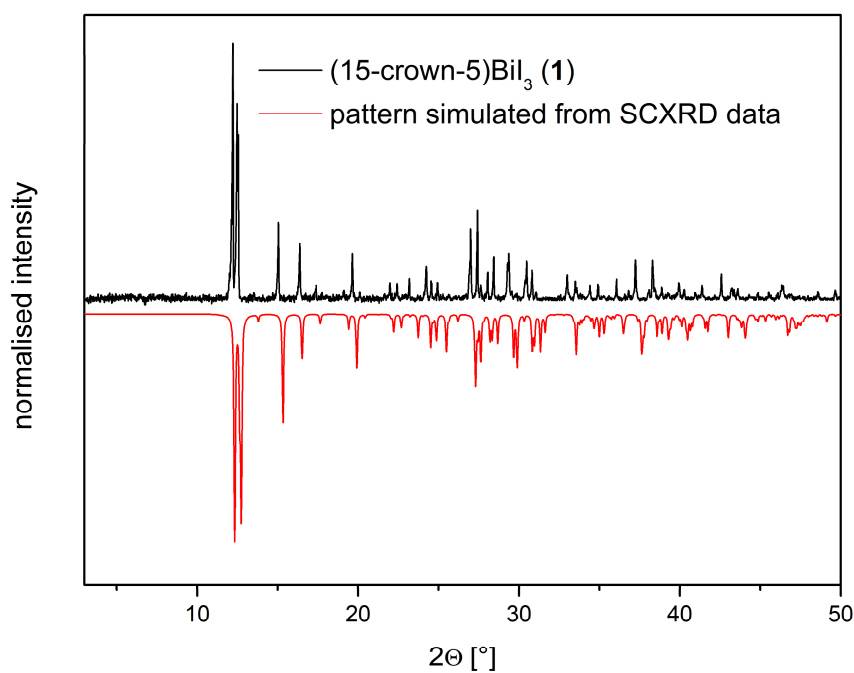

**Figure S3.** Powder diffraction pattern of (15-crown-5)BiI$_3$ (**1**).

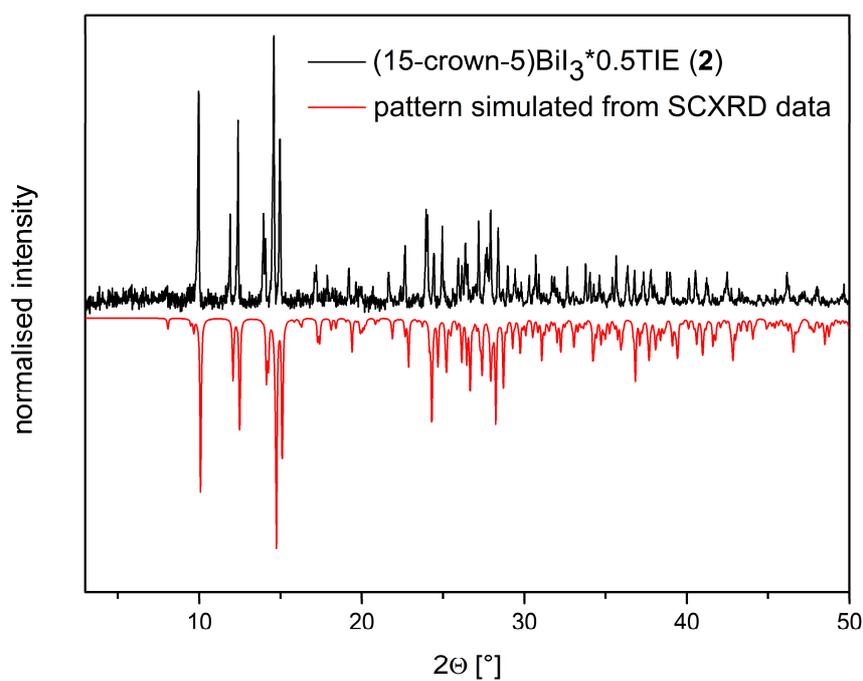

**Figure S4.** Powder diffraction pattern of (15-crown-5)BiI$_3$·0.5TIE (**2**).

# Elemental Analysis

**Micro-X-ray Fluorescence Spectroscopy (μ-XRF)**

The data were recorded on a *Bruker M4 Tornado*, equipped with an Rh-target X-ray tube and a Si drift detector. The emitted fluorescence photons were detected with an acquisition time of 180 s and 240 s. Upon deconvolution of the spectra, quantification of the elements was achieved based on the Bi L and I L radiation.

**Table S3:** μ-XRF analysis of **1**

|  | Atom % (exp) | Atom % (calc) | Element ratio (exp) | Element ratio (calc) |
|---|---|---|---|---|
| Bi (L-series) | 32.25 | 25.00 | 1 | 1 |
| I (L-series) | 67.75 | 75.00 | 2.1 | 3 |

**Table S4:** μ-XRF analysis of **2**

|  | Atom % (exp) | Atom % (calc) | Element ratio (exp) | Element ratio (calc) |
|---|---|---|---|---|
| Bi (L-Serie) | 18.98 | 16.67 | 1 | 1 |
| I (L-Serie) | 81.02 | 83.33 | 4.3 | 5 |

**Energy-Dispersive X-Ray Spectroscopy (EDX) and Scanning Electron Microscopy (SEM)**

Measurements were performed on a *JEOL JIB-4601F* with a *XFlash 5010* EDX detector with 129 eV resolution using the *Bruker Esprit 2.1* software package.

**Table S5:** EDX analysis of **1**

|  | Atom % (exp) | Atom % (calc) | Element ratio (exp) | Element ratio (calc) |
|---|---|---|---|---|
| Bi | 24.89 | 25.00 | 1 | 1 |
| I | 75.11 | 75.00 | 3.03 | 3 |

**Table S6:** EDX analysis of **2**

|  | Atom % (exp) | Atom % (calc) | Element ratio (exp) | Element ratio (calc) |
|---|---|---|---|---|
| Bi | 16.15 | 16.67 | 1.00 | 1 |
| I | 83.85 | 83.33 | 5.20 | 5 |

## Thermal Analysis

The thermal behavior was studied by TGA/DSC. The experiments were carried out on a *NETSCH STA 409 C/CD* with a heating rate of 10 °C/min in a constant flow of 80 ml/min Ar.

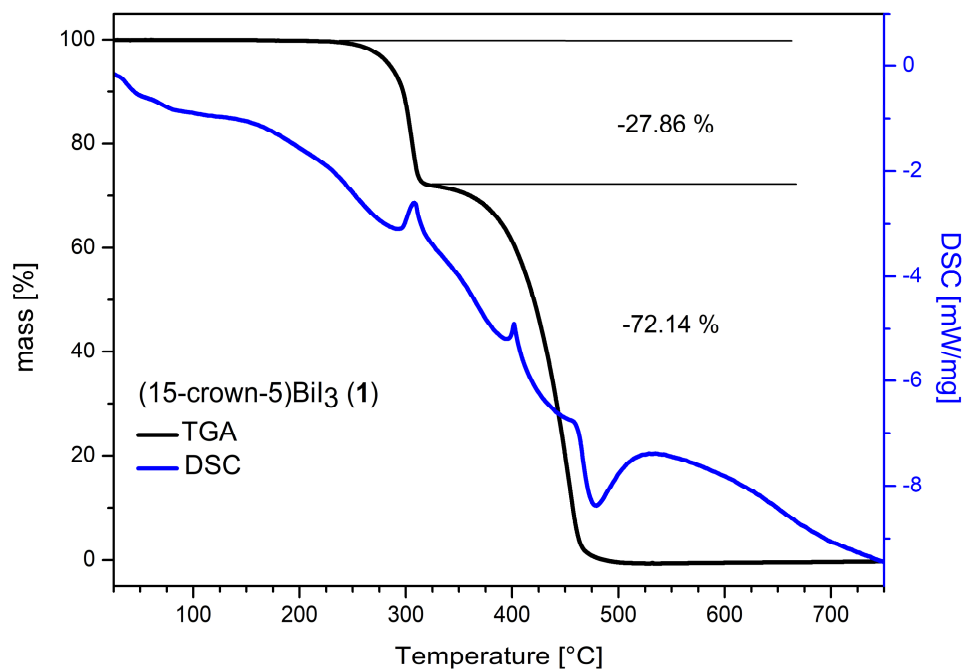

**Figure S5.** TGA/DSC of (15-crown-5)BiI$_3$ (**1**) (16.5 mg). A two-step decomposition process starts at 260°C with the loss of the crown ether and continues at 300°C with the sublimation of BiI$_3$.

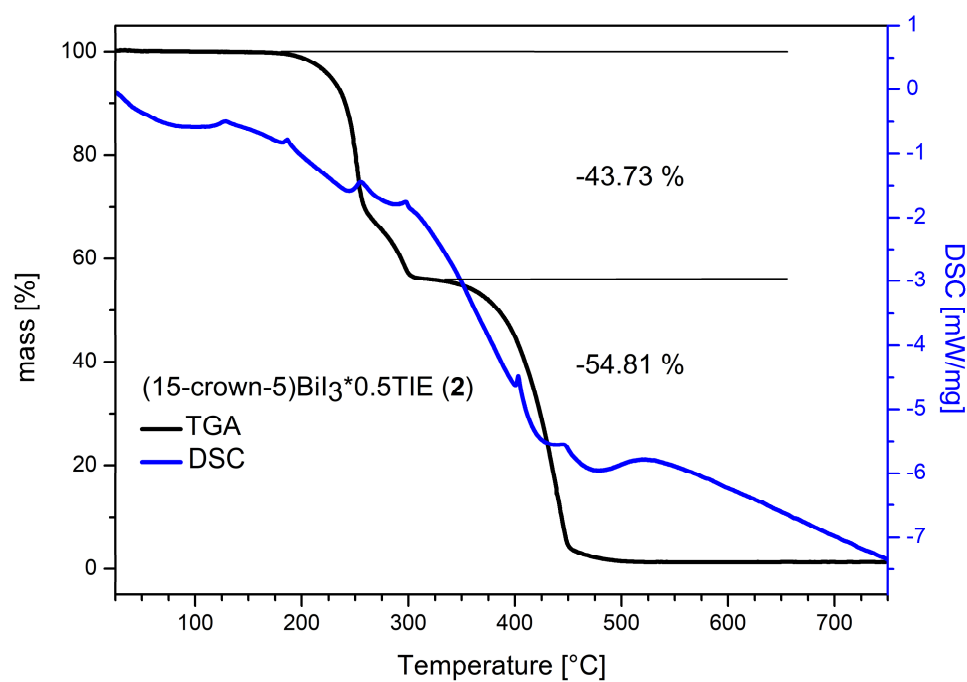

**Figure S6.** TGA/DSC of (15-crown-5)BiI$_3$·0.5TIE (**2**) (18.9 mg). A multi-step decomposition process starts at 200°C with the concomitant decomposition of TIE and the crown ether and continues at 300°C with the sublimation of BiI$_3$.

## Optical properties

Optical absorption spectra were recorded on a *Varian Cary 5000* UV/Vis/NIR spectrometer in diffuse reflectance mode employing a Praying Mantis accessory (*Harrick*). For ease of viewing, raw data was transformed from % Reflectance R to Absorbance A according to A = log (1/R).[2]

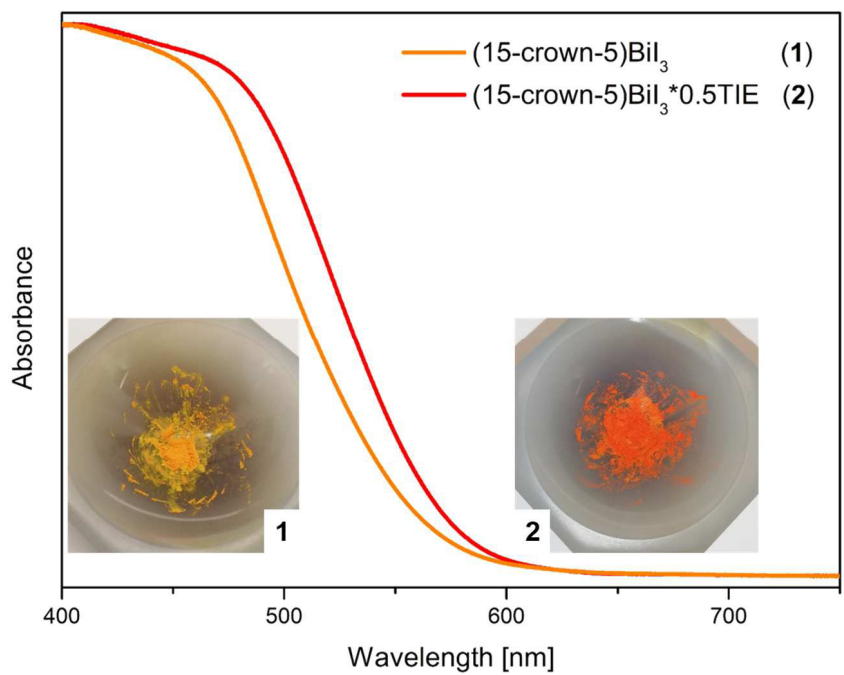

**Figure S7.** Optical absorption spectra of (15-crown-5)BiI$_3$ (**1**) and (15-crown-5)BiI$_3$·0.5TIE (**2**), measured in diffuse reflectance in the range of 400 nm to 800 nm. Photographs of the compounds are shown as insets. Optical band gaps estimated from the absorption spectrum are 2.23 eV (555 nm) (**1**) and 2.16 eV (575 nm) (**2**).

## Vibrational spectroscopy

IR spectra were recorded on a *Bruker Tensor 37* FT-IR spectrometer equipped with an ATR-Platinum measuring unit. Raman data were collected on an *S&I MonoVista CRS+* device. The measurements were performed with a laser wavelength of 488 nm (TIE) and 532 nm and a grating of 1800 grooves per mm. The measurement had duration of 10 s and 10 co-additions.

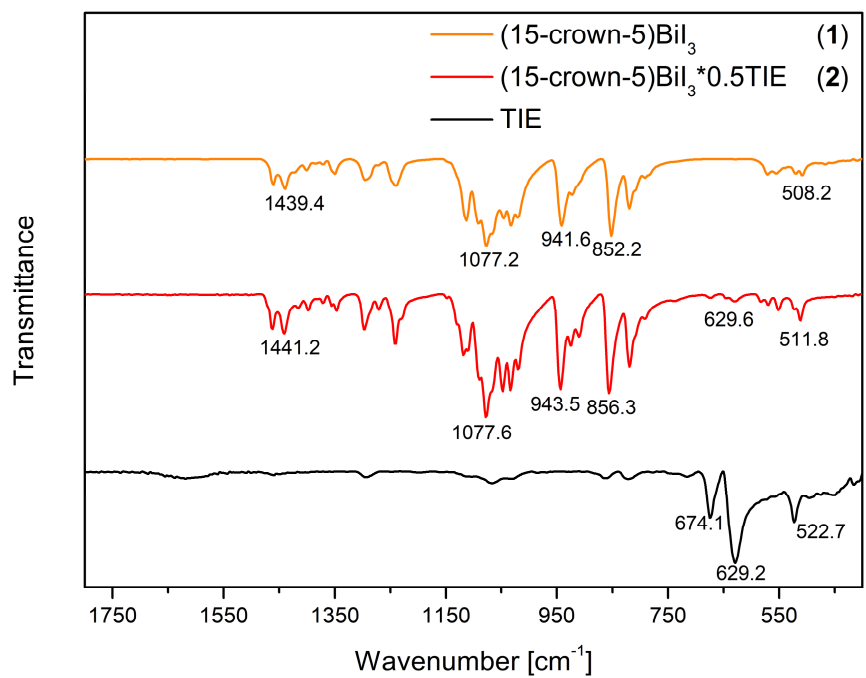

**Figure S8.** Comparison of IR-spectra. The strongest IR bands of TIE, C–I stretching occur at 522.7, 629.2 and 674.1 cm$^{-1}$, similar to literature reports.[3] In the spectrum of **2**, these bands are much weaker, possibly due to a stronger inequivalence of the C-I bonds, and partially superimposed by bands of the (15-crown-5)BiI$_3$ moiety, precluding a direct comparison.

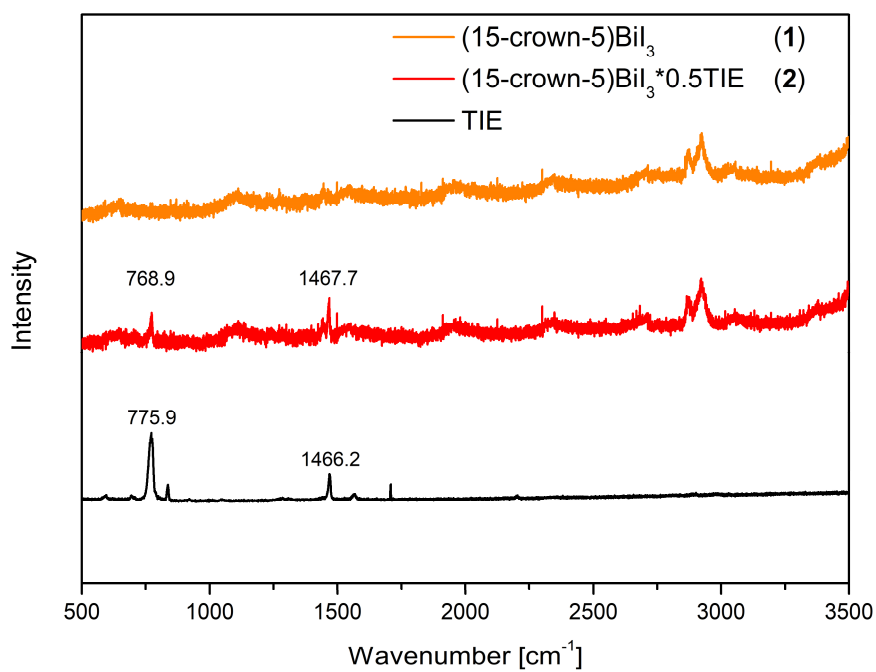

**Figure S9.** Comparison of Raman-spectra. The C–I stretching band observed at 775.9 cm$^{-1}$ in TIE is red-shifted to 768.9 cm$^{-1}$ in **2**, a similar change as has been observed in other halogen bonded TIE adducts.[4]

## Crystal Photographs

Pictures were taken using a *Zeiss SteREO Discovery.V8* optical microscope.

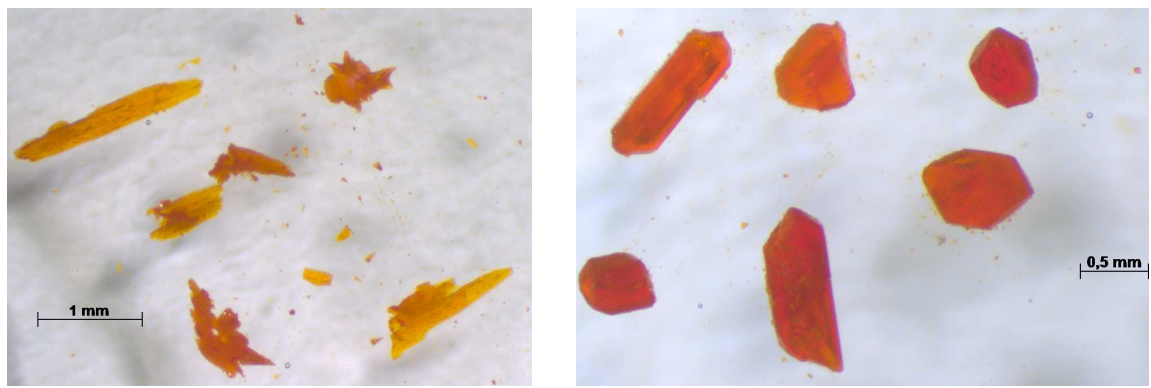

**Figure S10.** Photographs of crystals of **1** (left) and **2** (right).